\documentclass[11pt]{article}
\usepackage{epsfig,dsfont}

\begin{document}
\sffamily

\thispagestyle{empty}
\vspace*{15mm}

\begin{center}

{\LARGE  New overlap construction of Weyl fermions 
\vskip2mm
on the lattice}

\vskip20mm
Christof Gattringer and Markus Pak
\vskip5mm
Institut f\"ur Physik, Unversit\"at Graz \\
8010 Graz, Austria 
\end{center}
\vskip30mm

\begin{abstract}
In a recent article Hasenfratz and von Allmen have suggested a fixed point
action for two flavors of Weyl fermions on the lattice with gauge 
group SU(2). The block-spin
transformation they use maps the chiral and vector symmetries of the
underlying vector theory onto two equations of the Ginsparg-Wilson 
(GW) type. We show that an overlap Dirac operator can be constructed which 
solves both GW equations simultaneously. We discuss the
properties of this overlap operator and its projection onto lattice
Weyl fermions which seems to be free of artefacts, in particular the
projection operators are independent of the gauge field.  
\end{abstract}
\vskip30mm

\begin{center}
{\sl To appear in Nuclear Physics B.}
\end{center}

\setcounter{page}0
\newpage
\noindent
{\Large  Introduction}
\vskip3mm
\noindent
Lattice gauge theory made a big leap forward when it was understood how the
chiral symmetry of a massless vector-like theory manifests itself on the
lattice. The lattice Dirac operator has to obey the Ginsparg-Wilson (GW) 
equation
\cite{GiWi} and solutions of this equation are, e.g., the overlap operator
\cite{overlap} and fixed point fermions \cite{FP}. A lattice variant 
of the continuum 
chiral rotations was constructed and the axial anomaly identified in an
elegant way \cite{luscher}.

Immediately after the vector-like case was understood, the problem of a
lattice regularization of chiral gauge theories was attacked 
\cite{chiralgauge}. The chosen approach was to project a solution of the 
Ginsparg-Wilson equation for the vector-like theory onto its left-handed 
components. An unpleasant feature of this projection is that the left-acting 
projectors and their counterparts acting to the right have a different
structure. Due to the GW equation 
one of the two sets of projectors must depend on the Dirac operator
and thus on the gauge field. This leads to unwanted CP violating terms 
\cite{HaLatt02,HaBi} and 
the problem of having an additional gauge field dependence in the integration
measure of the lattice path integral of the chiral gauge theory. 

In a recent paper \cite{HaVA} Hasenfratz and von Allmen revisited the 
problem of transferring continuum symmetries onto the lattice using block spin
transformations. In particular they
analyzed the connection between anomalies and the blocking prescription used
for mapping the continuum theory onto the lattice. The key insight is that
the blocking kernel must break all anomalous symmetries of the target theory. 
For a 
two-flavor vector-like SU(2) gauge theory in a symmetric representation of the
fermion action, they proposed a particular blocking kernel with the correct
symmetry breaking pattern. A corresponding fixed point action was derived
which obeys two Ginsparg-Wilson type equations for the flavor singlet chiral
and vector symmetries of the underlying vector theory. The theory can be
projected to left-handed components, and for the resulting
theory of Weyl fermions the anomalies were shown to be correctly mapped onto the
lattice. 

In this paper we use the two Ginsparg-Wilson equations from \cite{HaVA} as
a starting point. We show that it is possible to construct a simultaneous 
overlap solution for both equations. The physical and doubler branches are
studied and we establish that the correct continuum limit is approached in the
physical sector and the doublers decouple. We discuss different
representations of the overlap solution, in particular as the sign function of a
hermitian matrix. The overlap operator may be projected to Weyl fermions in a
symmetric way without generating artefacts and the measure in the path
integral assumes a simple form. 

\vskip6mm
\noindent
{\Large  Continuum theory and its symmetries on the lattice}
\vskip3mm
\noindent
In our construction we start from a vector theory which we later project 
to a chiral gauge theory. In the continuum the action has the form
\begin{eqnarray}
\label{contact}
S[\overline{\psi},\psi] & \; = \; & 
\int \! d^4x \, \overline{\psi} \,
\gamma_\mu ( \, \overrightarrow{\partial}_{\!\mu} + i A_\mu \,) \psi 
\\
& \; = \; & 
\frac{1}{2} \int \! d^4x \, 
\Big[ \, \overline{\psi} \, \gamma_\mu 
( \, \overrightarrow{\partial}_{\!\mu} + i A_\mu\, ) \psi
- \psi^T \gamma_\mu^T (\, \overleftarrow{\partial}_{\!\mu} + i A^T_\mu \, ) 
\overline{\psi}^T \, \Big] \; .
\nonumber
\end{eqnarray}
$\overline{\psi}$ and $\psi$ are Grassmann valued  Dirac spinors that carry
SU(2) color and SU(2) flavor indices which we  suppress at the moment but will
make them explicit later in the lattice Dirac operators we consider.  The
gauge field assumes values in su(2), the algebra of our gauge group SU(2).

In the second line of (\ref{contact}) we have already identically rewritten
the action in a  form that is suitable for identifying the symmetric 
representation
of the fermions which we will use for our lattice discretization. The
superscript $T$ denotes transposition.  Introducing new 8-component fermion
spinors $\Psi^T = (\psi^T, \overline{\psi})$  we can write the fermion action
as 
\begin{equation}
S[\Psi] = \frac{1}{2}  
\int \! d^4x \, \Psi^T  D^{cont}  \Psi \;\;\; , \;\;\; 
D^{cont} =  \left[\! \begin{array}{cc}
0 &\! - \gamma_\mu^T 
( \, \overleftarrow{\partial}_{\!\mu}\! +\!  i A^T_\mu \,) \\
\! \gamma_\mu ( \, \overrightarrow{\partial}_{\!\mu} \! +\! i A_\mu\,) & 0
\end{array} \!\right] \!.
\label{contdirac}
\end{equation}
The continuum Dirac operator $D^{cont}$ has a structure of $4\times4$ blocks
which act on the upper and lower four components of our new spinors $\Psi$. In
flavor space $D^{cont}$ is diagonal.

The Dirac operator contains also the transpose gauge field $A_\mu^T$ and for
later use we note a relation for this transposition. The su(2)-valued
gauge fields can be expressed in terms of the Pauli matrices  $\sigma_1,
\sigma_2, \sigma_3,$
\begin{equation}
A_\mu = A_\mu^{(1)} \frac{\sigma_1}{2} + A_\mu^{(2)}  \frac{\sigma_2}{2} + 
A_\mu^{(3)}  \frac{\sigma_3}{2}  \quad , \quad 
A_\mu^T = A_\mu^{(1)}  \frac{\sigma_1}{2} - A_\mu^{(2)}  \frac{\sigma_2}{2} + 
A_\mu^{(3)}  \frac{\sigma_3}{2} \; ,
\end{equation}
and since only $\sigma_2$ is anti-symmetric, while $\sigma_1$ and $\sigma_3$
are symmetric, transposition of the gauge field corresponds to flipping the
sign of the second component. Obviously we find
\begin{equation}
A_\mu^T \; = \; - \sigma_2 \, A_\mu \, \sigma_2 \; = \; 
\varepsilon^c \, A_\mu \, \varepsilon^c \qquad
\mbox{with} \quad \varepsilon^c \; = \; i \sigma_2 \; .
\label{atransposition}
\end{equation}
The superscript $c$ attached to the $\varepsilon$-tensor refers to the color
indices, which  $\varepsilon^c$ acts on.

Having outlined the details of our continuum target theory, we now need to
discuss how symmetries of the continuum theory manifest themselves on the
lattice. In particular we are interested in the flavor singlet chiral and vector
transformations of the action (\ref{contact}) which in the notation with the
8-component spinors $\Psi$ are generated by 
\begin{equation}
\Gamma_5 =  \left[\! \begin{array}{cc}
\mathds{1}^c \! \otimes \! \gamma_5 \! \otimes \! \mathds{1}^f &  0 \\
0 & \mathds{1}^c \! \otimes \! \gamma_5 \! \otimes \! \mathds{1}^f
\end{array} \!\right] 
, \; \;
\Gamma_V = \left[\! \begin{array}{cc}
\mathds{1}^c \! \otimes \! \mathds{1}^d \! \otimes \! \mathds{1}^f &  0 \\
0 & - \mathds{1}^c \! \otimes \! \mathds{1}^d \! \otimes \! \mathds{1}^f
\end{array} \!\right].
\end{equation}
Here we now have made explicit the action on all involved indices, color, Dirac
and flavor, and use the
superscripts  $c,d,f$ to denote the action in color-, Dirac- and flavor-space.
For the upper and lower components of our 8-spinors we will continue to use
vector/matrix notation for notational convenience. The corresponding
transformations of the 8-component spinors read 
\begin{equation}
\Psi \; \longrightarrow \; \exp(i \epsilon \Gamma_5) \, \Psi \qquad , \qquad 
\Psi \; \longrightarrow \; \exp(i \epsilon \Gamma_V) \, \Psi \; .
\label{contfieldtrafo}
\end{equation}
The action (\ref{contdirac}) is invariant since $D^{cont}$ anti-commutes
both with  $\Gamma_5$ and $\Gamma_V$.

The lattice manifestation of continuum symmetries is most conveniently
obtained by analyzing the symmetries of a block spin transformation,
\begin{equation}
e^{-\frac{1}{2} \Phi^T D \Phi} \; = \; \int \! {\cal D}[\Psi] \, 
e^{-(\Phi - \Psi^B)^T E^{-1} (\Phi - \Psi^B)} \, e^{-S[\Psi]} \; .
\label{blocktrafo}
\end{equation}
Here $\Psi^B$ denotes a blocked field living on the lattice, which is obtained
from its continuum counterpart through a suitable gauge covariant blocking
prescription. The fields of the lattice theory are denoted by $\Phi$ and the
corresponding lattice Dirac operator is $D$. The blocked continuum fields
$\Psi^B$ and the lattice fields $\Phi$ are coupled through a blocking kernel
$E^{-1}$ in the first exponent on the rhs. The blocking kernel is diagonal in
the discrete lattice space-time indices and we use vector/matrix notation for
the lattice indices, both in the term that couples the blocked and lattice
fields, as well as for the quadratic form  $\Phi^T D \Phi$ in the lattice
action on the lhs. On the rhs.~the continuum field $\Psi$ is
integrated over in a path integral to eliminate the degrees of freedom above
the lattice cutoff.

Exploring the invariance of the continuum action it is possible to map the
continuum symmetries onto the lattice. They manifest themselves 
(see e.g.~\cite{GiWi,HaNiVA,GaPa} for a derivation) in
Ginsparg-Wilson equations for the lattice Dirac operator $D$, and since
here we are interested in both, the vector and the chiral symmetry, we obtain
two such equations ($a$ denotes the lattice spacing) 
\begin{eqnarray}
\Gamma_5 \, D & \, + \, & D \, \Gamma_5 \;\, = \;\, 
a \, D \, E \, \Gamma_5 \, D \; , 
\nonumber
\\
\Gamma_V \, D & \, + \, & D \, \Gamma_V \; = \; 
a \, D \, E \, \Gamma_V \, D \; .
\label{gwequations}
\end{eqnarray}
The continuum transformations (\ref{contfieldtrafo}) change to the 
corresponding symmetry transformations of the lattice fields (which we denote
only up to ${\cal O}(\epsilon)$),
\begin{equation}
\Phi \; \longrightarrow \, \Phi \, + \, i 
\epsilon \, \Gamma \, ( \mathds{1} - a E D/2 )\, \Phi 
\; + \; {\cal O}(\epsilon^2)
\quad , \quad
\Gamma \, = \, \Gamma_5 \, , \, \Gamma_V \; ,
\end{equation}
which change the integration measure of the lattice fields according to 
\begin{equation}
{\cal D}[\Psi] \; \longrightarrow \; {\cal D}[\psi] \; \Big( 1 + i \epsilon 
\mbox{Tr} \, [ \, \Gamma       
( \mathds{1} - a E D/2 ) \,  ] \; + \;  {\cal O}(\epsilon^2) \Big) \; ,
\end{equation}
a relation that may be used to identify the anomalies.

An important issue is the choice of the blocking kernel $E^{-1}$, which
couples the blocked and the lattice fields in the block-spin transformation
(\ref{blocktrafo}). The central construction principle  \cite{HaVA} is that
one must break all the symmetries which are anomalous in the quantized target
theory, while other, non-anomalous symmetries need not, but 
may be broken if it is
convenient to do so. Thus for the lattice version of a vector-like theory it
is sufficient to break the U(1) chiral symmetry, as was done in \cite{GiWi}
with a blocking matrix proportional to the identity.   Since here we are
ultimately interested in the chiral version of (\ref{contact}), e.g., the case
where we project to left-handed fermions, we need to break both the U(1)
chiral and the U(1) vector symmetries.  The choice suggested in \cite{HaVA}
reads
\begin{equation}
E \, =  \, i \left[\! \begin{array}{cc}
\varepsilon^c \otimes \overline{C} \otimes \varepsilon^f &  0 \\
0 & \varepsilon^c \otimes \overline{C} \otimes \varepsilon^f
\end{array} \!\right] \; .
\label{edef}
\end{equation}
The matrix $E$ commutes with both generators $\Gamma_5$ and $\Gamma_V$ and
thus has
the above discussed symmetry breaking pattern (a vanishing anti-commutator
would correspond to an unbroken symmetry). 

In $E$ tensors $\varepsilon^c$ and $\varepsilon^f$ for color and
flavor are used. Already introduced above, we here list some properties of
these $\varepsilon$-tensors which
we will need later,
\begin{equation}
\varepsilon \, = \, i \sigma_2 \quad , \quad \varepsilon \, = 
\, - \varepsilon^T \, = \, -
\varepsilon^\dagger \, = \, - \varepsilon^{-1} \; .
\label{epsilonproperties}
\end{equation}
In Dirac space $E$ applies a modified charge conjugation matrix
$\overline{C}$ which is related to the conventional charge conjugation matrix
$C$. We here use the chiral representation  where $C$ and $\overline{C}$ obey
\begin{eqnarray}
&& C \, = \, i \gamma_2 \gamma_4 
\quad , \quad 
\overline{C} \, = \, i \gamma_5 C 
\quad , \quad
\overline{C} C \, = \, C \overline{C} \, = \, i \gamma_5 \; ,
\label{cproperties}
\\
&& C \, = \, - C^T \, = \, C^\dagger \, = \, C^{-1} 
\quad , \quad
\overline{C} \, = \, - \overline{C}^{\,T} \, = \, - \overline{C}^{\,\dagger} 
\, = \, -\overline{C}^{\,-1} \; .
\nonumber 
\end{eqnarray}
Both, $C$ and $\overline{C}$ relate the $\gamma_\mu$ matrices to their
transpose,
\begin{equation}
\gamma_\mu^T \, = \, - C \gamma_\mu C  \quad , \quad 
\gamma_\mu^T \, = \, - \overline{C} \gamma_\mu \overline{C} \; .
\label{chargegamma}
\end{equation}
From (\ref{epsilonproperties}) and (\ref{cproperties}) follow the properties
of $E$,
\begin{equation}
E \, = \, - E^T \, = \, E^\dagger \, = \, E^{-1} 
\quad , \quad 
E\, \Gamma_5 = \Gamma_5 E 
\quad ,\quad 
E \, \Gamma_V = \Gamma_V E \; .
\label{eproperties}
\end{equation}

\vskip6mm
\noindent
{\Large  Overlap solution for the GW equations}
\vskip3mm
\noindent
We now show that a simultaneous overlap solution of the two Ginsparg-Wilson
equations  (\ref{gwequations}) can be constructed. 

We begin with examining the transposition properties of the SU(2) valued link
variables in the fundamental representation, which are used for introducing 
the gauge field on the lattice. An
element $U$ of SU(2) may be written as 
\begin{equation}
U \, = \, \exp\Big( i \vec{A} \cdot \vec{\sigma}/2\Big) \; = \; 
\mathds{1}^c \, \cos\Big((\vec{A}/2)^{\,2}\Big) \, + \, 
i \, \frac{\vec{A} \cdot \vec{\sigma}/2}{(\vec{A}/2)^2} \, 
\sin\Big((\vec{A}/2)^{\,2}\Big) \; ,
\end{equation}
with a real valued coefficient vector $\vec{A}$. Similar to the algebra valued
continuum field in (\ref{atransposition}), we may write the transpose of the
gauge link with the help of the $\varepsilon$-tensor 
$\varepsilon^c = i \sigma_2$,
\begin{equation}
U^T \; = \; \sigma_2 U^\dagger \sigma_2 \; = \; - 
\varepsilon^c U^\dagger \varepsilon^c \; .
\label{utransposition}
\end{equation}

The construction of the overlap solution makes use of a generalized Wilson Dirac
operator which is built from the naive discretization $V_\mu(x,y)$ of the
covariant derivative and the Wilson term  $S(x,y)$ which is proportional to
the covariant Laplace operator,
\begin{eqnarray}
&& V_\mu(x,y) \; = \; \frac{1}{2} \Big[ U_\mu(x) \,\delta_{x+\hat{\mu},y} \; - \; 
U_\mu(x-\hat{\mu})^\dagger \, \delta_{x-\hat{\mu},y} \Big]\; ,
\label{vsdefs}
\\
&& S(x,y) \; = \; 4 \, \mathds{1}^c \, \delta_{x,y} \; - \; 
\frac{1}{2} \sum_{\mu=1}^4 \Big[
U_\mu(x) \,\delta_{x+\hat{\mu},y} \; + \; 
U_\mu(x-\hat{\mu})^\dagger \, \delta_{x-\hat{\mu},y} \Big] \; .
\nonumber
\end{eqnarray}
The two terms, are anti-hermitian and hermitian, respectively, 
\begin{equation}
V_\mu^\dagger \; = \; - \, V_\mu \quad , \quad S^\dagger \; = \; S \; .
\end{equation}
Using the relation (\ref{utransposition}) their transposition properties may
be expressed as
\begin{equation}
V_\mu^T \; = \; \varepsilon^c \, V_\mu \, \varepsilon^c \quad , \quad 
S^T \; = \; - \, \varepsilon^c \, S \, \varepsilon^c \; .
\label{vstransposition}
\end{equation}

The generalized Wilson Dirac operator $D_W$ which we use for the overlap is 
constructed from $V_\mu$ and $S$, 
\begin{equation}
D_W \; = \; \frac{1}{a} \left[ \begin{array}{cc}
i \varepsilon^c S \otimes \overline{C} \otimes \varepsilon^f \; & \; 
-V_\mu^T \otimes \gamma_\mu^T \otimes \mathds{1}^f \\
V_\mu \otimes \gamma_\mu \otimes \mathds{1}^f \; & \;
i S \varepsilon^c \otimes \overline{C} \otimes \varepsilon^f 
\end{array} \right].
\label{dwdefinition}
\end{equation}
The derivatives $V_\mu$ are arranged such that in the naive continuum
limit they approach the continuum Dirac operator $D^{cont}$ as given in
(\ref{contdirac}). The term $S$, which removes the doublers, couples to $E$,
as can be seen by comparing the blocks on the diagonal in (\ref{dwdefinition})
to (\ref{edef}).

The overlap solution $D$ we present here is given by 
\begin{eqnarray}
D & \, = \, & \frac{1}{a} \, \Big[ \, E \; - \; A \, 
( \, E \, \Gamma_5 \, A \, E \, \Gamma_5 \, A \, )^{-1/2} \, \Big]
\nonumber \\ 
 & \, = \, & \frac{1}{a} \, \Big[ \, E \, - \, A \, 
( \, E \, \Gamma_V \, A \, E \, \Gamma_V \, A \, )^{-1/2} \, \Big] \; ,
\label{overlapdef} 
\end{eqnarray}
where
\begin{equation}
A \; = \;  E \, - \, a \, D_W  \; .
\end{equation}
In (\ref{overlapdef}) we have displayed the overlap operator in two different
forms.  The fact that the two forms are identical follows from the relation 
\begin{equation}
E \, \Gamma_5 \, A \, E \, \Gamma_5  \; = \; 
E \, \Gamma_V \, A \, E \, \Gamma_V \; ,
\end{equation}
which may be established using the identities 
(\ref{epsilonproperties}) -- (\ref{chargegamma}). 
Using for each of the two GW equations
(\ref{gwequations}) the suitable form of $D$, the  
equations (\ref{gwequations}) reduce to
\begin{eqnarray}
&& E \, \Gamma_5 \, A \, 
( \, E \, \Gamma_5 \, A \, E \, \Gamma_5 \, A)^{-1/2} \, 
E \, \Gamma_5 \, A \, ( \, E \, \Gamma_5 \, A \, E \, \Gamma_5 \, A)^{-1/2} 
\; = \; \mathds{1} \; , 
\label{giwimodified} 
\\
&& E \, \Gamma_V \, A \, 
( \, E \, \Gamma_V \, A \, E \, \Gamma_V \, A)^{-1/2} \, 
E \, \Gamma_V \, A \, ( \, E \, \Gamma_V \, A \, E \, \Gamma_V \, A)^{-1/2} 
\; = \; \mathds{1} \; .
\nonumber
\end{eqnarray}
These two identities hold trivially as can be seen
using the spectral theorem for the
inverse square root. This establishes that (\ref{overlapdef}) solves both
Ginsparg-Wilson equations (\ref{gwequations}). 

The Dirac operator may be brought into a second form, using the sign function
of a hermitian matrix. This is possible due to the fact that the products 
$E \, \Gamma_5 A$ and $E \, \Gamma_V A$ both are hermitian matrices, 
\begin{equation}
( \, E \, \Gamma_5 \, A \,)^\dagger \; = \; E \, \Gamma_5 \, A 
\quad , \quad 
( \, E \, \Gamma_V \, A \,)^\dagger \; = \; E \, \Gamma_V \, A \; .
\label{hermiticity}
\end{equation}
These equations may be established using
(\ref{epsilonproperties}) -- (\ref{chargegamma})
and (\ref{vstransposition}). From (\ref{hermiticity}) it
follows that the arguments of the inverse square roots in (\ref{overlapdef})
are squares of hermitian matrices. Using $(E \Gamma_5)^2 = \mathds{1}$ and 
$(E \Gamma_V)^2 = \mathds{1}$, we find the sign representation of the overlap 
operator
\begin{equation}
D \; = \; \frac{1}{a} \Big[ E \, - \, E \, \Gamma_5 \, \mbox{sign} \, 
( E \, \Gamma_5\, A ) 
\Big] \; = \; \frac{1}{a} 
\Big[ E \, - \, E \, \Gamma_V \, \mbox{sign} \, ( E \, \Gamma_V\, A ) \Big] \; .
\label{overlapsign}
\end{equation}

Having established, that the overlap operator (\ref{overlapdef}) solves the
two Ginsparg-Wilson equations (\ref{gwequations}), we still need to show that
it gives rise to the correct continuum limit and removes the doublers. We
begin this analysis with the physical branch where $V_\mu$ and $S$ behave as
\begin{eqnarray}
V_\mu & \, = \, & 
a \, \Big[ \, \overrightarrow{\partial}_{\!\mu} + i A_\mu \, \Big] \; + \; 
{\cal O}(a^2)  \; ,
\label{physbranch} \\
V_\mu^T & \, = \, &
a \, \Big[ \,  \overleftarrow{\partial}_{\!\mu} + i A^T_\mu \, \Big] \; + \;  
{\cal O}(a^2) \; ,
\nonumber \\
S  & \, \propto \, & {\cal O}(a^2) \; .
\nonumber
\end{eqnarray}
Inserting these into (\ref{dwdefinition}) and subsequently in
(\ref{overlapdef}), one finds with the help of 
(\ref{epsilonproperties}) -- (\ref{chargegamma}) and (\ref{vstransposition})   
that in the physical branch our overlap operator 
approaches the correct continuum operator $D^{cont}$ as given in  
(\ref{contdirac}),
\begin{equation}
D \; = \; D^{cont} \; + \; {\cal O}(a) \; .
\end{equation}
For the doubler branches one has
\begin{equation}
V_\mu \; \propto \; {\cal O}(a) \quad , \quad 
S \; = \; \mathds{1}^c \, 2k \; + \; {\cal O}(a^2) \; , \; k = 1,2,3,4 \; ,
\end{equation}
and inserting these again into (\ref{dwdefinition}) and (\ref{overlapdef})
gives rise to the behavior
\begin{equation}
D \; = \; \frac{2}{a} \, E \; + \; {\cal O}(1) \; .
\end{equation}
The rhs.~diverges as $a \rightarrow 0$ and thus the doublers decouple. It is
interesting to note, that the term $S$, which removes the doublers, 
couples to the
blocking matrix $E$, which has eigenvalues $+1$ and $-1$. Consequently the
doubler modes end up symmetrically at the positions $\pm 2/a$ in the complex
plane. This is different from the usual vector-like overlap operator, where
the doublers all end up on the positive real axis near $2/a$. 
We stress, however, that
this is not a peculiarity of the overlap solution given here. Also the fixed
point solution of \cite{HaVA}, which for the free case can be computed in
closed form following \cite{BiWi}, distributes the doublers symmetrically.  

Finally it is interesting to observe, that when going back to the
conventional notation with 4-spinors, the term that removes the doublers
assumes the form
\begin{equation}
i \, \frac{2}{a} \, \Big[ \,
\psi^T \epsilon^c \otimes \overline{C} \otimes \epsilon^f \psi \, + \,
\overline{\psi} \, \epsilon^c \otimes \overline{C} \otimes \epsilon^f \,
\overline{\psi}^T \, \Big] \; .
\label{doublerterm}
\end{equation}
The same structure is obtained for the free fixed point operator 
computed by direct blocking from
the continuum. We stress that such a term cannot be formulated within
the usual bilinear representation of the fermion action, and is possible only
in the symmetric form used here.

\newpage
\noindent
{\Large Properties of $D$ and its projection to Weyl fermions}
\vskip3mm
\noindent
In this section we now discuss some of the key properties of $D$ which are
necessary for the discussion \cite{HaVA} of the anomalies of our theory and
its projection to Weyl fermions. 

The analysis of the anomaly given in \cite{HaVA} makes use of the two
GW equations (\ref{gwequations}) and the fact that the Dirac
operator $D$ is $\widehat{\Gamma}_5$-hermitian, i.e., it obeys
\begin{equation}
\widehat{\Gamma}_5  \, D \widehat{\Gamma}_5 \; = \; D^\dagger \; ,
\label{g5hermiticity}
\end{equation}
where
\begin{equation}
\widehat{\Gamma}_5  \; = \; \left[\! \begin{array}{cc}
0 & \mathds{1}^c \! \otimes \! \gamma_5 \! \otimes \! \mathds{1}^f \\
\mathds{1}^c \! \otimes \! \gamma_5 \! \otimes \! \mathds{1}^f & 0
\end{array} \!\right] \; .
\label{g5hat}
\end{equation}
For our overlap operator Eq.~(\ref{g5hermiticity})  
can be shown by noting that 
\begin{equation}
\widehat{\Gamma}_5  \, A \, \widehat{\Gamma}_5 \; = \; A^\dagger
\; = \; E \, \Gamma_5  \, A \, E \, \Gamma_5  \; ,
\end{equation}
where the last identity is a consequence of (\ref{eproperties}) and 
(\ref{hermiticity}). Thus the overlap Dirac operator can also be written as
\begin{equation}
D \; = \; \frac{1}{a} \, \Big[ \, E \; - \; A \, 
\Big( \, \widehat{\Gamma}_5 \, A \, \widehat{\Gamma}_5 \, A \, \Big)^{-1/2} 
\, \Big] \; = \; \frac{1}{a} \,
\Big[ \, E \, - \, \widehat{\Gamma}_5 \, 
\mbox{sign} \, \Big( \widehat{\Gamma}_5 \, A \Big) \, \Big]\; .
\end{equation}
The $\widehat{\Gamma}_5$-hermiticity of $D$ 
then follows from this equation together
with the $\widehat{\Gamma}_5$-hermiticity of $A$ and 
$\widehat{\Gamma}_5^{\;2} = \mathds{1}$.

For a Dirac operator that obeys (\ref{gwequations}) and (\ref{g5hermiticity}) 
it was shown in \cite{HaVA} that the correct anomaly structure of a
vector-like theory emerges, i.e.,
the vector transformation is free of anomalies, while for the flavor singlet
chiral transformation the correct axial anomaly is found. 

Having fully understood the vector-like theory on the lattice, we can now turn
to the chiral gauge theory. For our symmetric representation of the fermions,
we define suitable left- ($P_-$) and right-handed ($P_+$) projectors,
\begin{equation}
P_\pm \; = \;  \left[\! \begin{array}{cc}
\mathds{1}^c \! \otimes \! \frac{\mathds{1}^d \pm \gamma_5}{2} \! 
\otimes \! \mathds{1}^f &  0 \\
0 & \mathds{1}^c \! \otimes \! \frac{\mathds{1}^d \mp \gamma_5}{2}  \! 
\otimes \! \mathds{1}^f 
\end{array} \!\right] \; .
\end{equation}
The projectors obey
\begin{equation}
P_\pm \, P_\pm \, = \, P_\pm \; , \; P_\pm \, P_\mp \, = \, 0 \; , \; 
P_+ \, + P_- \, = \, \mathds{1} \; , \; P_\pm^T \, = \, P_\pm \; ,
\end{equation}
which are the usual properties of projectors and their symmetry under 
transposition which we stress
explicitly as this is important for the symmetric fermion representation used 
here.
 
The Dirac operator $D_-$ for left-handed Weyl fermions is then obtained by
projecting the vector-like operator $D$ from, e.g, the right,
\begin{equation}
D_- \; = \; D \, P_- \; = \; P_- \, D \; .
\label{Dminus}
\end{equation} 
In this equation we have already noted that the projection may also be done
from the left (with the same projector). The fact that $D$ and the projectors
commute is an important consistency relation which follows from the 
possibility to write the projectors as
\begin{equation}
P_\pm \; = \; \frac{1}{2} \, \Big[ \mathds{1} \pm \Gamma_V \Gamma_5 \Big] \; ,
\end{equation}
together with the two GW equations (\ref{gwequations}).
We finally remark that also the Weyl operator $D_-$ obeys the GW
equations, as is expected from the corresponding symmetry in the continuum 
\cite{GaPa}. 

For the projected operator $D_-$ it was shown in \cite{HaVA} that the correct
fermion number anomaly is obtained. The arguments are again only based on the
GW equations (\ref{gwequations}) and the $ \widehat{\Gamma}_5$-hermiticity 
(\ref{g5hermiticity}). Since our overlap operator obeys all of these, the proof
of \cite{HaVA} applies and we
conclude that the projected overlap operator has the correct chiral anomaly. 
At the same time the projection operators are independent of $D$ and thus
no unwanted gauge field dependence is introduced in the path integral 
measure of the chiral theory.

\vskip10mm
\noindent
{\bf Acknowledgments:}
We thank Peter Hasenfratz for interesting discussions.
This work is partly supported by the FWF project P20330-N16.

\end{document}